\documentclass{crtbt} 

\DocReference{http://crtbt.grenoble.cnrs.fr/helio/ \\submitted to Physics of Fluids}
\PSLogo{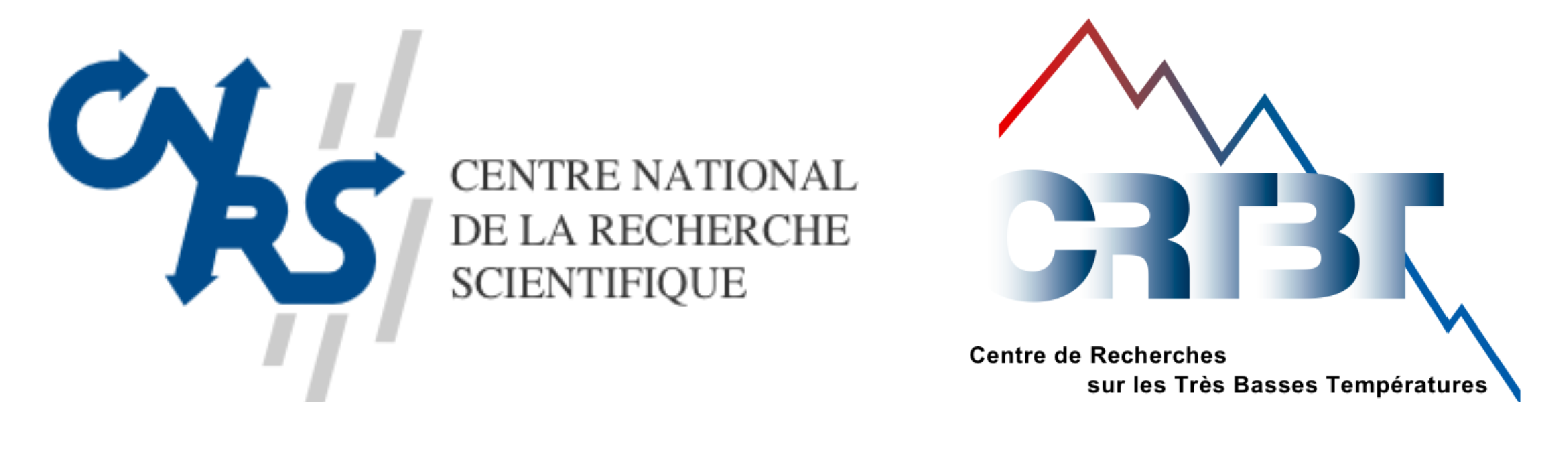}
\LogoHeight{1.8cm}
    
\Author{P.-E. Roche, F. Gauthier, B. Chabaud and B. H\'ebral} 
\Title{Ultimate Regime of Convection: \\ robustness to poor thermal properties of the plates\\} 

\Distribution{ccsd and CondMat}

\begin{document} 
\Maketitle 

\Summary{
A transition to Kraichnan ultimate regime of convection has been reported in very high Rayleigh numbers experiments, but not in all of them. These apparently contradictory results can be explained by a recent phenomenological model  which accounts for the non-ideality of the plate thermal properties [Chill\`a et al, Physics of Fluids 16:2452 (2004)]. In this paper, we present a direct test of this model, using a low conductivity plate. We found an unaltered transition, not compatible with the model's predictions.
}

\section{Motivation}

In 1962, Robert Kraichnan predicted \cite{Kraichnan1962} that a fully-turbulent regime of convection will settle in Rayleigh-B\'enard cells for an intense thermal driving force (very high Rayleigh number $Ra$). This regime is characterized by turbulent boundary layers and consequently an enhanced heat transport efficiency (Nusselt number $Nu$) compared to the other convection regimes. Kraichnan predicted a $Nu\sim Ra^{1/2}$ dependence, with a logarithmic correction which reduces the effective exponent right above the transition to this so-called ultimate regime.

Only 5 experiments and one numerical simulation have explored Rayleigh numbers at least at $Ra=10^{13}$ and higher.  For all of them, the Prandtl number was in the intermediate region ($0.7<Pr<20$). The experiments have been conducted nearly in the same conditions : cylindrical cells of aspect ratio 0.5 or 1, cryogenic helium as the working fluid,  copper top and bottom plates and thin stainless steel side wall. These very high Rayleigh number studies can be separated into two sets according to the measured $Nu(Ra)$ dependence. For two experiments \cite{Wu, Niemela2000}, this dependence is smooth and follows approximatively a 0.3 exponent power law up to $Ra\sim 10^{17}$ for one of them\cite{Niemela2000}, while for two other experiments \cite{Chavanne1997,Niemela2003} and the simulation \cite{Kenjeres2002}, the exponent increases from 0.3 to approximately 0.4 in the $Ra=10^{11}-10^{12}$ region. This second regime has been interpreted as Kraichnan ultimate regime and this interpretation was strengthened by the measurement of the full 0.5 exponent in a rough surfaces cell \cite{RochePRE2001}.

No obvious parameter -such as the Prandtl number or the cell aspect ratio- seems to be correlated with the occurrence of the transition among these few studies. Therefore, understanding what can favor or suppress occurence of the ultimate regime remains today a major challenge for the understanding of very high Rayleigh number convection \cite{Kadanoff2001,SommeriaPREFACE}.

Recently, Chill\`a and colleagues proposed a phenomenological model \cite{Chilla2004} trying to explain these apparently contradictory results. This model asseses the non-ideality of the plates' thermal properties : if the conductivity and/or the heat capacity of the plates are not high enough, a constant temperature cannot be sustained at the fluid-plate interface during the development of thermal plumes. As a consequence, the global heat transfer is shown to saturate below Kraichnan prediction. The Chilla-rastello condition ($Cr$) to observe the transition is :

\begin{equation}
Cr>0.8
\end{equation}

Where $Cr$ is a plate quality factor which depends on the ratios of plate to fluid conductivities $\lambda _{p} \over \lambda _{f}$, heights $a \over h$ ($a$ is the plate thickness) and heat capactities, as well as the usual dimensionless parameters $Nu$, $Pr$, aspect ratio $\Gamma$ and Reynolds number $Re$. For the cryogenic experiments previously mentionned, a thin plate and quasistatic approximations can be applied to the general expression of $Cr$. One gets \cite{Chilla2004} :

\begin{equation}
Cr \simeq \frac{\pi ^2}{\Gamma ^2} \frac{a}{h} \frac{\lambda _{p}}{\lambda _{f}}\frac{1}{Re Pr}
\end{equation}

For reference, we should mention two other recent studies of finite plate conductivity effect on convection \cite{Verzicco_plate,Hunt2003}, although they are not directly related to the ultimate regime issue. Some previous works are reported in these two papers.

In this paper, we present two experiments that have been conducted to test Chill\`a \textit{et al.} model. We took out of the shelves two 20-cm high Rayleigh-B\'enard cells in which the transition have been observed in the past: one with smooth surfaces \cite{Chavanne1997} and the other one with rough (corrugated) surfaces \cite{RochePRE2001}. Both bottom plates were removed and replaced with two others, geometrically-similar but made out of brass: one with a smooth surface and the other with a rough one. The corresponding $Cr$ parameter for these new bottom plates is smaller than $0.1$ for $Ra>10^{11}$. According to Chill\`a \textit{et al.}, the global heat transfer $Nu(Ra)$ in the ultimate regime should be significantly reduced compared to the original all-copper experiments.

\section{Experimental set-up and results}

In order to make the comparison with the all-copper experiments straightforward, the experimental set-up and the operating procedure are very similar to those of these previous studies \cite{Chavanne2001, RochePRE2001}. Since this experimental information is detailed in the previous papers, we shall only report here the differences with these experiments.

\begin{figure}
\centerline{\includegraphics[width=10cm]{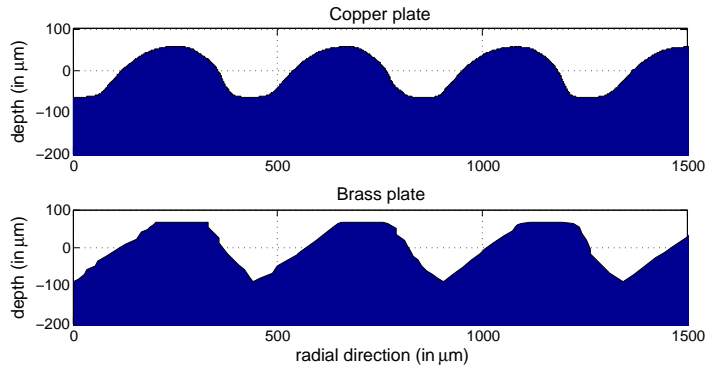}}
\caption{Profilometer measurement of the roughness profile along a bottom plate diameter. Distances are given in microns. Upper plot : copper plate, Lower plot : brass plate.
}
\label{fig:grooves}
\end{figure}

The conduction of the brass plates has been measured with the inductive technique already used to measure the conductivity of the copper plates \cite{RocheTHESE}. We found $k_{Brass}=4.0\,Wm^{-1}K^{-1}$ $\pm10\%$ at 4.2\,K, to be compared with $k_{Cu}=1090\,Wm^{-1}K^{-1}$ and $880\,Wm^{-1}K^{-1}$ at 4.2\,K for the smooth surfaces and rough surfaces copper plates. In our experimental conditions, the heat capacity of brass is comparable to the one of copper \cite{Conte}. The surface roughness of the corrugated plate has been measured with a profilometer. The grooves are 450\,$\mu$m apart \cite{erratumRBR}, 145\,$\mu$m deep and look like the reference copper plate roughness (see fig. \ref{fig:grooves}). Both 25\,mm-thick brass bottom plates are heated from below. In order to minimize the heater footprint on the other side of the plates, the heating wire has been uniformly and carefully varnished with a 1.5\,mm spacing on the bottom surface.

A special attention has been dedicated to the temperature measurements on the axis of the brass plates, near the fluid-brass interface. Finite elements simulations \cite{mefisto} have been used to optimize the shape of the copper thermometer holders. Optimization was performed for $10^9<Ra<10^{14}$ and by modeling the fluid's thermal boundary layer as a diffusive and spatially uniform slab of thickness $h/2Nu$, where $h$ is the height of the cell. The thermometer holders are inserted into the brass plates, in long $\phi$\,4\,mm holes ending with a 3\,mm long $\phi$\,3\,mm thread, itself ending 2\,mm away from the surface in contact with the fluid. 
The holes and the thermometer holders are responsible for a local non-uniformity of temperature at the fluid-brass interface. According to the simulation, and for all the data presented in this paper, the magnitude of this effect is less than 1\% of the temperature difference across the fluid. To this extend, the temperature measurement can be assumed as non-invasive. A correction needs to be applied to compensate for the temperature drop between the thermometer holders and the fluid-brass interface. According to the simulation, this correction represents a 0.11\,K/W resistance, with negligible $Nu$ dependence, which corresponds to a effective position of the probe 3.5mm from the fluid-brass interface. For all the data presented in this paper, this correction represents less than 20\% of the temperature difference across the fluid (less than 10\% for the smooth surfaces cell).

Temperature on both the top and bottom plates have been measured with doped Ge resistors, in-situ calibrated  : 1) with respect to each other, 2) with respect to a commercially calibrated one \cite{Lakeshore}, 3) with respect to the fluid critical point. The estimated uncertainty on the temperature difference between the top and bottom plate is 1\,mK. All the data presented have been obtained with a temperature difference across the fluid larger than 20\,mK.

The arbitrarily Boussinesq criterion is that no fluid parameter should vary by more than $20\%$ across cell. Recent improvements in the knowledge of helium fluid properties \cite{RocheJLTP2004} have been taken into account in the calculation of $Ra$ and $Nu$. $Nu$ has been corrected for the side wall conduction using the analytical correction formula presented in  \cite{RocheEPJB2001}. The magnitude of this correction is nearly the same for all the cells. It decreases very rapidly with $Ra$ and represents less than a $5\%$ correction above $Ra=10^{12}$. 

\begin{figure}
\centerline{\includegraphics[width=10cm]{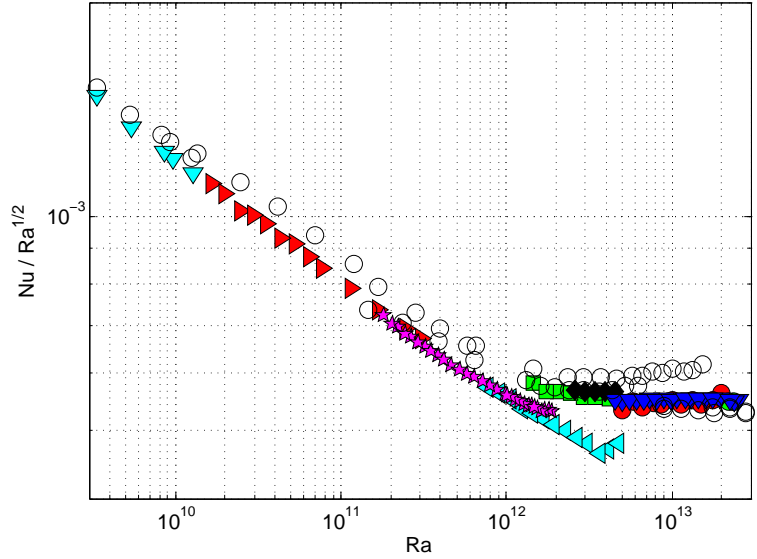}}
\caption{Compensated Nusselt number $Nu/Ra^{1/2}$ versus $Ra$ for the rough surfaces experiments. Open circles : copper bottom plate, Solid symbols : brass bottom plate.
}
\label{fig:rough}
\end{figure}

\begin{figure}
\centerline{\includegraphics[width=10cm]{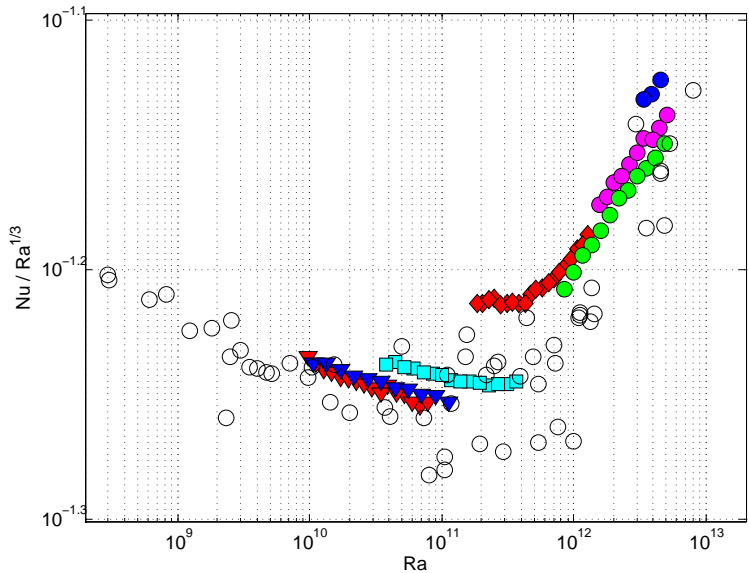}}
\caption{Compensated Nusselt number $Nu/Ra^{1/3}$ versus $Ra$ for the smooth surfaces experiments. Open circles : copper bottom plate, Solid symbols : brass bottom plate.
}
\label{fig:smooth}
\end{figure}

Figure \ref{fig:rough} and \ref{fig:smooth} shows the heat transfer data in the rough and smooth surfaces ``brass'' experiments. The various solid symbols are associated with data acquired for a fixed cell density and a nearly fixed average temperature in the cell. The Prandtl number associated with each data set is in the $0.7<Pr<5.4$ range. For comparison, the all-copper experiment data, restricted to the same $Pr$ window, are also plotted (open circles).

On Figure~\ref{fig:rough}, the rough surfaces $Nu$ data are compensated by $Ra^{1/2}$ so that Kraichnan's $Nu(Ra)$ power law corresponds to a plateau.  Such plateaux are found for both the brass and the all-copper experiments, and -within experimental uncertainty- they fall on each other. Some extra data between $Ra=3.10^{13}$ and $Ra=6.10^{13}$ are not shown because they don't meet our various criteria (Boussinesq, Prandtl window, temperature-drop maximum correction and minimum temperature difference across the fluid). Nevertheless these extra data for the two experiments keep falling on each other.

On Figure~\ref{fig:smooth}, the smooth surfaces $Nu$ data are now compensated by $Ra^{1/3}$. To our knowledge, the $1/3$ exponent is the highest exponent predicted by the theories of turbulent convection, let aside Kraichnan's regime. Above $Ra=4.10^{11}$, if we fit our data with a local power law, we find an effective exponent of $d\log (Nu) / d\log (Ra) \sim0.4$, significantly larger than the $1/3$ prediction. The scattering of the data from the reference all-copper experiment is larger because the temperature difference across the cell hasn't been average as long. Here again, within the experimental uncertainty, the data from both the brass and the all-copper experiments fall on each other.

\section{Concluding remarks}

The two experiments presented here are intended to test Chill\`a \textit{et al.} model \cite{Chilla2004}, which was able to explain why the ultimate regime is observed in some experiments but not all of them. Within experimental uncertainties, we find that the substitution of a copper plate by its replica made out of brass has no consequence on the global heat transfer, in particular within the ultimate regime. This result is in clear contradiction with the model.
 
It seems difficult to adjust the model parameters while preserving its ability to discriminate between the cryogenic experiments \cite{Wu,Chavanne1997,Niemela2000,RochePRE2001,Niemela2003}. This suggests that the critical difference between these experiments has to be searched elsewhere. A candidate is the large scale flow structure (LSFS). Its bimodality, which was evidenced \cite{RocheEPL2002,Verzicco2003} for $Ra<10^{11}$ may extend to higher $Ra$ as suggested by the two figures of this paper. In 2001, it was conjectured \cite{RocheTHESE} that this transition to the ultimate regime could only occur for one LSFS, at least below $Ra \sim 10^{17}$.
 
\section{Acknowledgment}
We are indebted to Bernard Castaing and Francesca Chill\`a for their various inputs in this work and to Fran\c{c}ois Debray and Christophe Trophime for assistance with finite element simulations. This work was supported by the R\'egion Rh\^one-Alpes, under contract number 301491302.

\bibliographystyle{plain}


\begin{thebibliography}{10}


\bibitem{Kraichnan1962}
R.~Kraichnan.
\newblock Turbulent thermal convection at arbitrary {P}randtl numbers.
\newblock {\em Phys. Fluids}, 5:1374, 1962.

\bibitem{Wu}
X.Z.~Wu and A.~Libchaber
\newblock {\em Scaling relations in thermal turbulence: The aspect-ratio dependence}.
\newblock {\em Phys. Rev. A}, 45:842, 1992.

\bibitem{Niemela2000}
J.J. Niemela, L.~Skrbek, K.R. Sreenivasan, and R.J. Donnelly.
\newblock Turbulent convection at very high {R}ayleigh numbers.
\newblock {\em Nature}, 404:837--840, 2000.

\bibitem{Chavanne1997}
X.~Chavanne, F.~Chill\`a, B.~Castaing, B.~H\'ebral, B.~Chabaud, and J.~Chaussy.
\newblock Observation of the ultimate regime in {R}ayleigh-{B}\'enard
  convection.
\newblock {\em Phys. Rev. Lett.}, 79:3648--3651, 1997.

\bibitem{Niemela2003}
J.J. Niemela, L.~Skrbek, K.R. Sreenivasan, and R.J. Donnelly.
\newblock Confined turbulent convection.
\newblock {\em J. Fluid Mech.}, 481:355--84, 2003.

\bibitem{Kenjeres2002}
S.~Kenjere\v{s} and K.~Hanjali\`c.
\newblock Numerical insight into flow structure in ultraturbulent thermal
  convection.
\newblock {\em Phys. Rev. E}, 66:036307, 2002.

\bibitem{RochePRE2001}
P.-E. Roche, B.~Castaing, B.~Chabaud, and B.~H{\'e}bral.
\newblock Observation of the 1/2 power law in {R}ayleigh-b{\'e}nard convection.
\newblock {\em Phys. Rev. E}, 63:045303(R) 1--4, 2001.

\bibitem{Kadanoff2001}
L.~P. Kadanoff.
\newblock Turbulent heat flow: Structures and scaling.
\newblock {\em Phys. Today}, 54:34--39, 2001.

\bibitem{SommeriaPREFACE}
J.~Sommeria.
\newblock The elusive ``ultimate state'' of thermal convection.
\newblock {\em Nature}, 398:294, 1999.

\bibitem{Chilla2004}
F.~Chill\`a, M.~Rastello, S.~Chaumat, and B.~Castaing.
\newblock Ultimate regime in {R}ayleigh {B}\'enard convection : the role of
  plates.
\newblock {\em Physics of Fluids}, 16:2452--2456, 2004.

\bibitem{Verzicco_plate}
R.~Verzicco.
\newblock Effects of nonperfect thermal sources in turbulent thermal
  convection.
\newblock {\em Phys. Fluids}, 16:1965, 2003.

\bibitem{Hunt2003}
J.C.R. Hunt, A.J. Vrieling, A.J. Nieuwstadt, and H.J.S. Fernando.
\newblock Influence of the thermal diffusivity of the lower boundary on eddy
  motion in convection.
\newblock {\em J. Fluid Mech.}, 491:183, 2003.

\bibitem{Chavanne2001}
X.~Chavanne, F.~Chill\`a, B.~Chabaud, B.~Castaing and B.~H\'ebral.
\newblock Turbulent Rayleigh-B\'enard convection in gaseous and liquid He.
\newblock {\em Phys. Fluids}, 13:1300-20, 2001.

\bibitem{RocheTHESE}
P.-E. Roche
\newblock Convection thermique turbulente en cellule de {R}ayleigh-{B}\'enard cryog\'enique
\newblock PhD thesis, Universit\'e Joseph Fourier, Grenoble, 2001.
\newblock http://tel.ccsd.cnrs.fr/documents/archives0/00/00/18/94/

\bibitem{Conte}
R. R. Conte
\newblock {\em El\'ements de cryog\'enie}.
\newblock Masson et Cie Editeurs, 1970.

\bibitem{erratumRBR}
\newblock The distance between grooves previously reported in the copper rough cell was wrong by a factor two. This doesn't change the interpretation and conclusion of this previous paper which only relies on the grooves' thickness.

\bibitem{mefisto}
The \textit{mefisto} finite element package has been used.
\newblock It is available at http://www.ann.jussieu.fr/\~{ }perronnet/mefisto.gene.html
\newblock Some of the results have been cross-checked with a commercial finite element software (ANSYS).

\bibitem{Lakeshore}
Model GR-200B-2500
\newblock {\em Lake Shore Cryotronics Inc, Ohio}.

\bibitem{RocheJLTP2004}
P.-E. Roche, B.~Castaing, B.~Chabaud, and B.~H{\'e}bral.
\newblock Heat transfer in turbulent {R}ayleigh-{B}\'enard convection below the
  ultimate regime.
\newblock {\em J. Low Temp. Phys.}, 134:1011--1042,
  2004.

\bibitem{RocheEPJB2001}
P.-E. Roche, B.~Castaing, B.~Chabaud, B.~H{\'e}bral, and J.~Sommeria.
\newblock Side wall effects in {R}ayleigh b{\'e}nard experiments.
\newblock {\em Eur. Phys. J. B}, 24:405--408, 2001.
\newblock This rough surfaces cell was built to test a consequence of Kraichnan theory stating that in such a cell the logarithmic correction can be ``frozen'' and the $1/2$ power law measured.

\bibitem{RocheEPL2002}
P.-E. Roche, B.~Castaing, B.~Chabaud, and B.~H{\'e}bral.
\newblock {P}randtl and {R}ayleigh numbers dependences in {R}ayleigh {B}\'enard convection
\newblock {\em Europhys. Lett.}, 58:693--698, 2002.

\bibitem{Verzicco2003}
R.~Verzicco and R.~ Camussi.
\newblock Numerical experiments on strongly turbulent thermal convection in a slender cylindrical cell.
\newblock {\em J. Fluid Mech.}, 477:19-49, 2003.



\end{thebibliography}

\end{document}